\documentclass[aps,prev,preprintnumbers,floatfix,nofootinbib, superscriptaddress]{revtex4-2}
\pdfoutput=1
\usepackage[total={6.5in,9in},top=1in,headsep=0.1in,headheight=2in]{geometry}
\usepackage{amsmath}
\usepackage{amsfonts}
\usepackage{amssymb}
\usepackage{graphicx}
\usepackage{bm}
\usepackage{times}
\usepackage{hyperref}
\usepackage{slashed}
\usepackage{color}

\newcommand\brazilreport{
\begin{center}
  \rule[-0.2in]{\hsize}{0.01in}\\
  \rule{\hsize}{0.01in}\\
  \vskip 0.1in
  Submitted to the Proceedings of the II Strategy Plan for High Energy, Cosmology and Astroparticle Physics (HECAP)\\ 
 
  \rule{\hsize}{0.01in}\\
  \rule[+0.2in]{\hsize}{0.01in}\\[-2em]
\end{center}
}

\usepackage[firstpage=true]{background}
\backgroundsetup{contents={\parbox{6.5in}{\brazilreport}}, scale=1,placement=top,opacity=1,color=black,position={3.25in,1.2in}}

\usepackage{fancyhdr}
\fancypagestyle{plain}{%
  \fancyhf{}%
  \fancyhead[C]{}
  \fancyfoot[C]{\thepage}
}

\fancypagestyle{empty}{%
  \fancyhf{}%
  \fancyhead[C]{{\it Report of the Brazilian Community on Dark Matter 2024}}
  \fancyfoot[C]{\thepage}
}
\vspace{2cm}
\begin{document}

\title{\LARGE Brazilian Report on Dark Matter 2024}

%
%

\author{I. F. M. Albuquerque}
\affiliation{Instituto de F\'isica, Universidade de S\~ao Paulo, Brazil}

\author{J. Alcaniz}
\affiliation{Observat\'orio Nacional, 20921-400, Rio de Janeiro, Brazil}

\author{A. Alves}
\affiliation{Departamento de F\'isica, Universidade Federal de S\~ao Paulo, UNIFESP, Diadema, 09972-270, Brazil}


\author{J. Amaral}
\affiliation{Faculdade de Engenharia, Universidade do Estado do Rio de Janeiro, 20950-000, Rio de Janeiro, RJ, Brazil}

\author{C. Bonifazi}
\affiliation{Instituto de F\'isica, Universidade Federal do Rio de Janeiro (UFRJ),Caixa Postal 68528, CEP 21941-972, Rio de Janeiro, RJ, Brazil}
\affiliation{%
ICAS \& Instituto de Ciencias Físicas, ECyT-UNSAM \& CONICET, Buenos Aires, Argentina \\
}%
\author{H. A. Borges}
\affiliation{Instituto de F\'isica, Universidade Federal da Bahia, 40210-340, Salvador, BA, Brasil}

\author{S. Carneiro}
\affiliation{Instituto de F\'isica, Universidade Federal da Bahia, 40210-340, Salvador, BA, Brasil}
\affiliation{Observat\'orio Nacional, 20921-400, Rio de Janeiro, Brazil}

\author{L. Casarini}
\affiliation{Departamento de F\'isica, Universidade Federal de Sergipe, 49100-000, São Cristovão, SE, Brasil}

\author{D. Cogollo}
\affiliation{Departamento de F\'isica, Universidade Federal de Campina Grande,
Campina Grande, PB, Brazil}

\author{A. G. Dias}
\affiliation{Centro de Ci\^encias Naturais e Humanas, Universidade Federal do ABC, 09210-580, Santo Andr\'e-SP, Brazil}

\author{G. C. Dorsch}
\affiliation{Universidade  Federal  de  Minas  Gerais,  ICEx,  Dep.   de  F\'isica ,Av. Ant\^ onio  Carlos,  6627,  Belo  Horizonte,  MG,  Brasil,  CEP  31270-901}

\author{A. Esmaili}
\affiliation{Departamento de F\'isica, Pontif\'icia Universidade Cat\'olica do Rio de Janeiro, Rio de Janeiro 22452-970, Brazil}

\author{G. Gil da Silveira}
\affiliation{Instituto de F\'{\i}sica, Universidade Federal do Rio Grande do Sul, Av. Bento Gon\c{c}alves, 9500\\
Porto Alegre, Rio Grande do Sul, CEP 91501-970,
Brazil}
\affiliation{Departamento de F\'{\i}sica Nuclear e de Altas Energias, Universidade do Estado do Rio de Janeiro\\
CEP 20550-013, Rio de Janeiro, RJ, Brazil}

\author{C. Gobel}
\affiliation{Departamento de F\'isica, Pontif\'icia Universidade Cat\'olica do Rio de Janeiro, Rio de Janeiro 22452-970, Brazil}

\author{V. P. { Gon\c{c}alves}}
\affiliation{Institute of Physics and Mathematics, Federal University of Pelotas, \\
  Postal Code 354,  96010-900, Pelotas, RS, Brazil}

\author{A. S. Jesus}
\affiliation{Instituto de F\'isica, Universidade de S\~ao Paulo, Brazil}

\author{D. Hadjimichef}
\affiliation{Departamento de F\'{\i}sica Nuclear e de Altas Energias, Universidade do Estado do Rio de Janeiro\\ CEP 20550-013, Rio de Janeiro, RJ, Brazil}

\author{P. C. de Holanda}
\affiliation{Universidade Estadual de Campinas, Campinas, SP 13083-970, Brazil}

\author{R.F.L. Holanda}
\affiliation{Departamento de F\'{\i}sica, Universidade Federal do Rio Grande do Norte, 59078-970, Natal, RN, Brasil}

\author{E. Kemp}
\affiliation{Instituto de Física Gleb Wataghin, Universidade Estadual de Campinas, Campinas, SP 13083-859, Campinas, SP, Brazil}

\author{A. Lessa}
\affiliation{Centro de Ci\^encias Naturais e Humanas, Universidade Federal do ABC, 09210-580, Santo Andr\'e-SP, Brazil}

\author{A. Machado}
\affiliation{Universidade Estadual de Campinas, Campinas, SP 13083-970, Brazil}

\author{M.V T. Machado}
\affiliation{Instituto de F\'{\i}sica, Universidade Federal do Rio Grande do Sul, Av. Bento Gon\c{c}alves, 9500\\
Porto Alegre, Rio Grande do Sul, CEP 91501-970,
Brazil}

\author{M. Makler}
\affiliation{Centro Brasileiro de Pesquisas F\'isicas, Rio de Janeiro, RJ, Brazil}
\affiliation{%
ICAS \& Instituto de Ciencias Físicas, ECyT-UNSAM \& CONICET, Buenos Aires, Argentina \\
}%
\author{V. Marra}
\affiliation{Departamento de Física \& Núcleo Cosmo-Ufes, Universidade Federal do Espírito Santo,
29075-910 Vitória, ES, Brazil}

\author{M. S. Mateus Junior}
\affiliation{Instituto de F\'{\i}sica, Universidade Federal do Rio Grande do Sul, Av. Bento Gon\c{c}alves, 9500\\
Porto Alegre, Rio Grande do Sul, CEP 91501-970,
Brazil}

\author{R. D. Matheus}
\affiliation{Instituto de F\'isica Te\'orica, Universidade Estadual Paulista, SP, Brazil}

\author{P. G. Mercadante}
\affiliation{Centro de Ci\^encias Naturais e Humanas, Universidade Federal do ABC, 09210-580, Santo Andr\'e-SP, Brazil}

\author{A. A. Nepomuceno}
\affiliation{Departamento de Ciências da Natureza, Universidade Federal Fluminense, 28895-532, Rio das Ostras, RJ, Brazil}

\author{R.M.P. Neves}
\affiliation{Universidade Estadual do Cear\'a (UECE) - 
 Faculdade de Educação, Ci\^encias e Letras de Iguatu,Av. D\'ario Rabelo s/n, Iguatu — CE, 63.500-00 — Brazil}

\author{C. Nishi}
\affiliation{Centro de Ci\^encias Naturais e Humanas, Universidade Federal do ABC, 09210-580, Santo Andr\'e-SP, Brazil}

\author{Y.M. Oviedo-Torres}
\affiliation{Millennium Institute for Subatomic Physics at the High-Energy Frontier (SAPHIR) of ANID, Fernandez Concha 700, Santiago, Chile}
\affiliation{International Institute of Physics,  Universidade Federal do Rio Grande do Norte,Campus  Universit\'ario,  Lagoa  Nova,  Natal-RN  59078-970,  Brazil}

\author{N. Pinto}
\affiliation{Centro Brasileiro de Pesquisas F\'isicas, Rio de Janeiro, RJ, Brazil}

\author{C. A. Pires}
\affiliation{Departamento de F\'isica, Universidade Federal de Campina Grande,
Campina Grande, PB, Brazil}

\author{E. Polycarpo}
\affiliation{Instituto de F\'isica, Universidade Federal do Rio de Janeiro (UFRJ),Caixa Postal 68528, CEP 21941-972, Rio de Janeiro, RJ, Brazil}

\author{F. S. Queiroz\footnote{Editor-Convener: Farinaldo S. Queiroz, e-mail: farinaldo.queiroz@ufrn.br}}
\affiliation{Departamento de F\'{\i}sica, Universidade Federal do Rio Grande do Norte, 59078-970, Natal, RN, Brasil}
\affiliation{International Institute of Physics,  Universidade Federal do Rio Grande do Norte,Campus  Universit\'ario,  Lagoa  Nova,  Natal-RN  59078-970,  Brazil}
\affiliation{Millennium Institute for Subatomic Physics at the High-Energy Frontier (SAPHIR) of ANID, Fernandez Concha 700, Santiago, Chile}

\author{T. Quirino}
\affiliation{Faculdade de Engenharia, Universidade do Estado do Rio de Janeiro, 20950-000, Rio de Janeiro, RJ, Brazil}

\author{M.~S.~Rangel}
\affiliation{Instituto de F\'isica, Universidade Federal do Rio de Janeiro (UFRJ),Caixa Postal 68528, CEP 21941-972, Rio de Janeiro, RJ, Brazil}

\author{P. Rebello Teles}
\affiliation{Centro Brasileiro de Pesquisas F\'isicas (CBPF), Rua Dr. Xavier Sigaud,
150 Urca, 22290-180, Rio de Janeiro, RJ, Brazil.}
\affiliation{CERN, EP Department 1211 Geneva, Switzerland}

\author{D. C. Rodrigues}
\affiliation{Departamento de Física \& Núcleo Cosmo-Ufes, Universidade Federal do Espírito Santo,
29075-910 Vitória, ES, Brazil}

\author{J. G. Rodrigues}
\affiliation{Observat\'orio Nacional, 20921-400, Rio de Janeiro, Brazil}

\author{P. S. Rodrigues da Silva}
\affiliation{Departamento de F\'isica, Universidade Federal de Campina Grande,
Campina Grande, PB, Brazil}

\author{R. Rosenfeld}
\affiliation{Instituto de F\'isica Te\'orica, Universidade Estadual Paulista, SP, Brazil}

\author{B. L. Sanchez-Vega}
\affiliation{Universidade  Federal  de  Minas  Gerais,  ICEx,  Dep.   de  F\'isica ,Av. Ant\^ onio  Carlos,  6627,  Belo  Horizonte,  MG,  Brasil,  CEP  31270-901}

\author{E. Segretto}
\affiliation{Universidade Estadual de Campinas, Campinas, SP 13083-970, Brazil}

\author{R. Silva}
\affiliation{Departamento de F\'{\i}sica, Universidade Federal do Rio Grande do Norte, 59078-970, Natal, RN, Brasil}
\affiliation{Departamento de F\'isica, Universidade do Estado do Rio Grande do Norte, Mossor\'o, 59610-210, Brasil}

\author{D. R. da Silva}
\affiliation{Centro Brasileiro de Pesquisas F\'isicas, Rio de Janeiro, RJ, Brazil}

\author{C. Siqueira}
\affiliation{Observatório Nacional, 20921-400, Rio de Janeiro - RJ, Brazil}
\affiliation{Departamento de F\'isica, Universidade Federal de Campina Grande,
Campina Grande, PB, Brazil}

\author{V. de Souza}
\affiliation{Instituto de F\'{\i}sica de S\~ao Carlos, Universidade de S\~ao Paulo.}

\author{S. Fonseca De Souza}
\affiliation{Universidade do Estado do Rio de Janeiro, Rio de Janeiro, Brazil}

\author{T. R. F. P. Tomei}
\affiliation{Núcleo de Computação Científica, Universidade Estadual Paulista, SP, Brazil}

\author{G. A. Valdiviesso}
\affiliation{Instituto de Ci\^encia e Tecnologia, Universidade Federal de Alfenas, Campus Po\c cos de Caldas, 37715-400, Po\c cos de Caldas - MG, Brazil}

\author{A. Viana}
\affiliation{Instituto de F\'{\i}sica de S\~ao Carlos, Universidade de S\~ao Paulo}

\author{Y. Villamizar}
\affiliation{Centro de Ci\^encias Naturais e Humanas, Universidade Federal do ABC, 09210-580, Santo Andr\'e-SP, Brazil}

\begin{abstract}
\vspace{5cm}
One of the key scientific objectives for the next decade is to uncover the nature of dark matter (DM). We should continue prioritizing targets such as weakly-interacting massive particles (WIMPs), Axions, and other low-mass dark matter candidates to improve our chances of achieving it. A varied and ongoing portfolio of experiments spanning different scales and detection methods is essential to maximize our chances of discovering its composition. This report paper provides an updated overview of the Brazilian community's activities in dark matter and dark sector physics over the past years with a view for the future. It underscores the ongoing need for financial support for Brazilian groups actively engaged in experimental research to sustain the Brazilian involvement in the global search for dark matter particles\footnote{We deeply mourn the passing of Professor Alex Dias, whose contributions to the field were invaluable. His dedication, insight, and passion left a lasting impact on colleagues, students, and the scientific community. He will be greatly missed, but his legacy will continue to inspire future generations.}.
\end{abstract}

\maketitle
\flushbottom

\section{Scientific Context}

\begin{figure}[h!]
    \centering
    \includegraphics[scale=0.75]{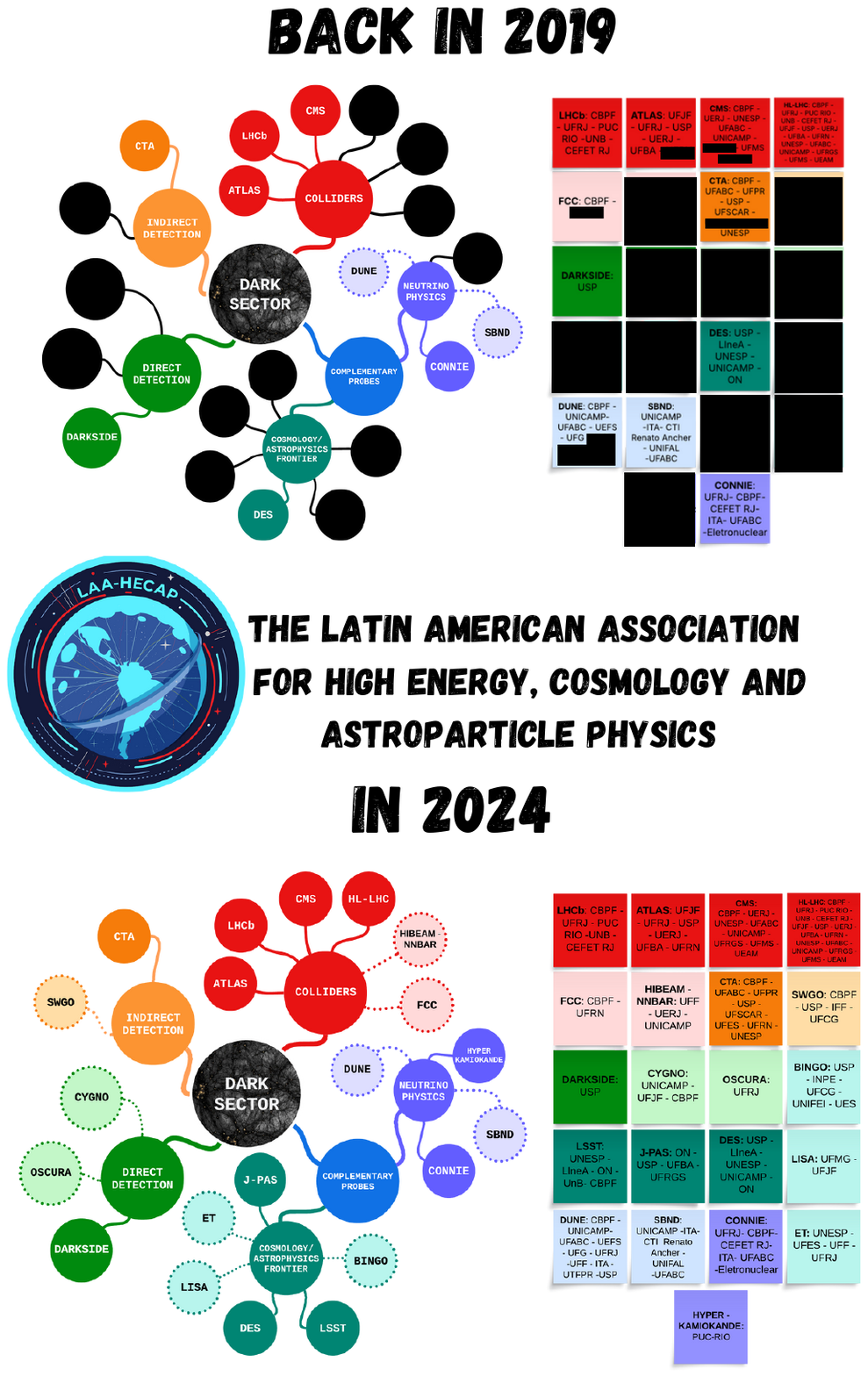}
    \caption{Map of the Brazilian involvement in dark matter and dark sector physics. Current detectors are delimited with solid lines and future detectors with dashed ones. Future detectors that passed the research and development phase and are already operating even at an early stage are delimited with solid lines, such as CTAO and LSST. We thank Lucia Angel, Ph.D. student from UFRN, for producing this beautiful figure.}
    \label{fig:dark_sectors}
\end{figure}

We have strong astronomical and cosmological evidence across many scales, ranging from velocities of stars in ultra-faint galaxies to the Cosmic Microwave Background, indicating that more than 80\% of the matter in our Universe consists of non-baryonic matter known as “Dark Matter” (DM). The steady collection of this evidence over the last eight decades suggests that DM consists of one or more new particles not contained within the Standard Model (SM) of Particle Physics. From these observations, we can set some basic requirements for DM. The dominant part of DM must be stable with a lifetime much longer than the age of the Universe,
it must form cosmological structures consistent with observations, and it must be produced in the early Universe. From these requirements, we can infer some of the general properties of DM. The formation of consistent cosmological structures sets a limit on the strength of DM interactions, both with the SM and
with itself. DM production in the early Universe may occur through a wide range of channels, but many of these require additional non-gravitational interactions in some form (See \cite{Arcadi:2024ukq} for an extensive review). Several scenarios for dark matter are consistent with these observations. In terms of elementary particles, masses from $10^{-22}$~eV all the way up to the Planck scale, in principle. In this report, we will focus on dark matter particle masses ranging from eV to 100 TeV and also discuss scenarios where dark matter is described by a fluid equation in cosmological scenarios. In FIG.~\ref{fig:dark_sectors} we exhibit the experiments with Brazilian involvement and the associated universities. It is clear that the study of the dark sector is a topic of common interest in the country, but we note that except for the UFCG involvement in BINGO and the UFRN participation in ATLAS and CTAO, the North-Northeastern universities do not directly participate in international experimental efforts.

Despite our ignorance concerning the mass of the dark matter particle, we have accumulated important information concerning dark matter, such that any dark matter candidate should fulfill some requirements: (i) it should yield the correct relic density; (ii) should be non-relativistic at matter-radiation equality to form structures in the early Universe in agreement with the observation; (iii) it should be effectively neutral otherwise it would form unobserved stable charged particles; (iv) and last but not least, it should be cosmological stable with a lifetime much larger than the age of the Universe to be consistent with cosmic rays and gamma-rays observations. Having these requirements in the back of our minds, the Brazilian community has been involved in several activities that will be classified according to the detection method below.

\section{Objectives }

The primary objective of this white paper is to detail the efforts of the Brazilian community in the search for dark matter and dark sector signals in the past four years \footnote{We will neither address the construction and operational costs nor computing requirements, as these aspects are best handled by the individual collaborations. Instead, this white paper will focus exclusively on the advancements in dark matter physics, rather than the financial challenges associated with these endeavors.}. We start addressing the usual dark matter complementarity, namely collider, direct and indirect detection. Later, we discuss complementary probes in terms of neutrino physics, cosmology, and astrophysics. We finish highlighting theoretical efforts that have sprouted recently and how they are aligned with the experiments. This white paper convincingly demonstrates that the community has both grown and consolidated, actively participating in large-scale and long-term international collaborations.

\section{Direct Detection}

Data from galaxy rotation curves shows the presence of a dark matter halo surrounding galaxies. Due to our relative motion to a stationary dark matter halo, we expect an incoming flux of dark particles in our detectors. Underground laboratories hope to observe low-energy nuclear or electron recoils rising from dark matter scatterings. By measuring the energy recoil, how this energy is deposited, and the shape of the scattering rate, one may discriminate dark matter signals from potential background sources through different readout techniques \cite{Billard:2021uyg}. The Brazilian involvement in direct dark matter detection has increased. Five years ago, it was limited to one researcher in the DarkSide collaboration. Now, we observe the presence of other researchers in projected direct detection experiments from different institutions. Increasing the number of physicists is important, given the paramount importance of direct detection in discovering dark matter. Below, we give a short description of the relevant detectors.\\

\paragraph{DarkSide}

\begin{figure}
    \centering
    \includegraphics[width=0.4\textwidth]{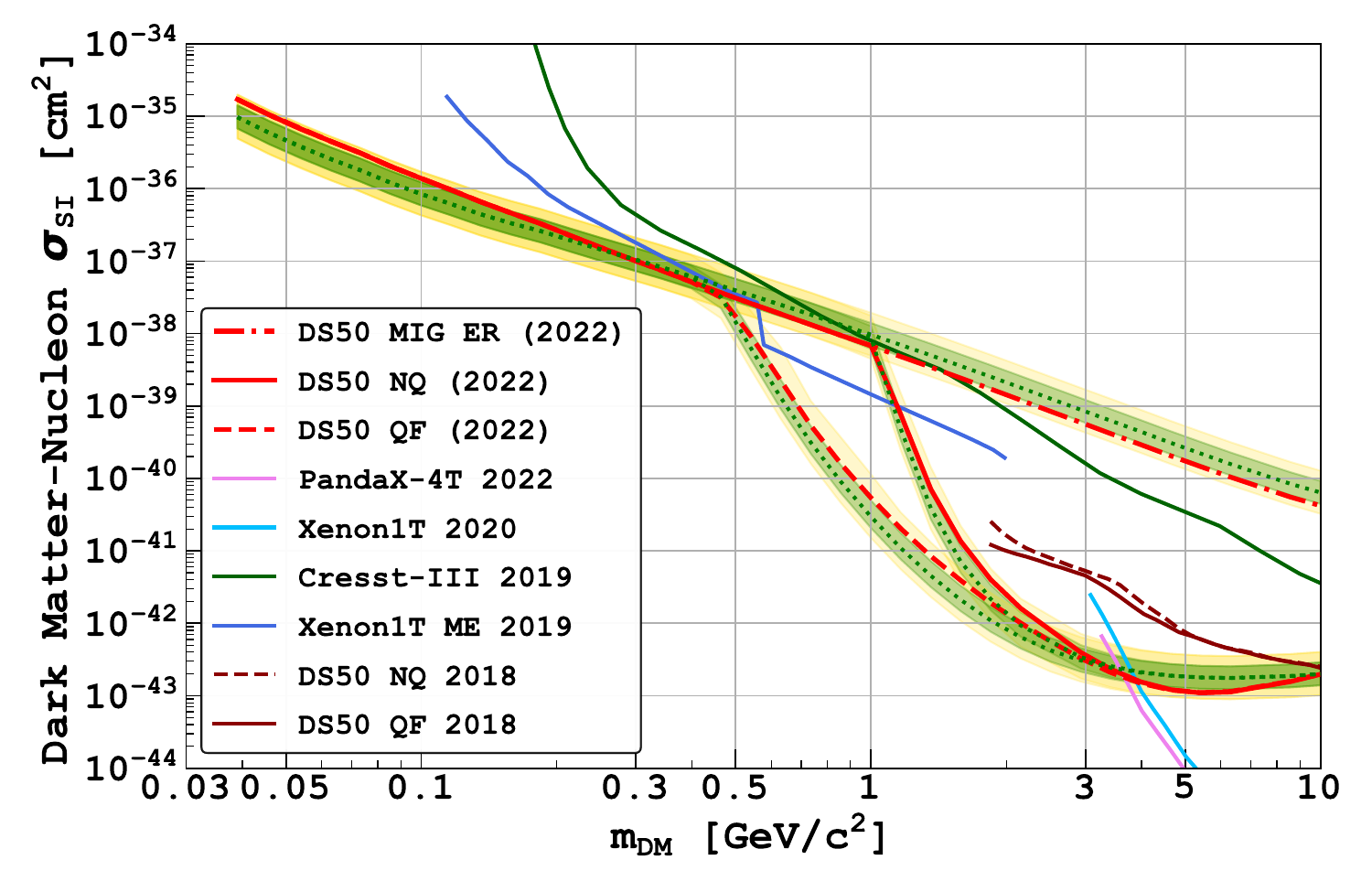}
    \includegraphics[width=0.4\textwidth]{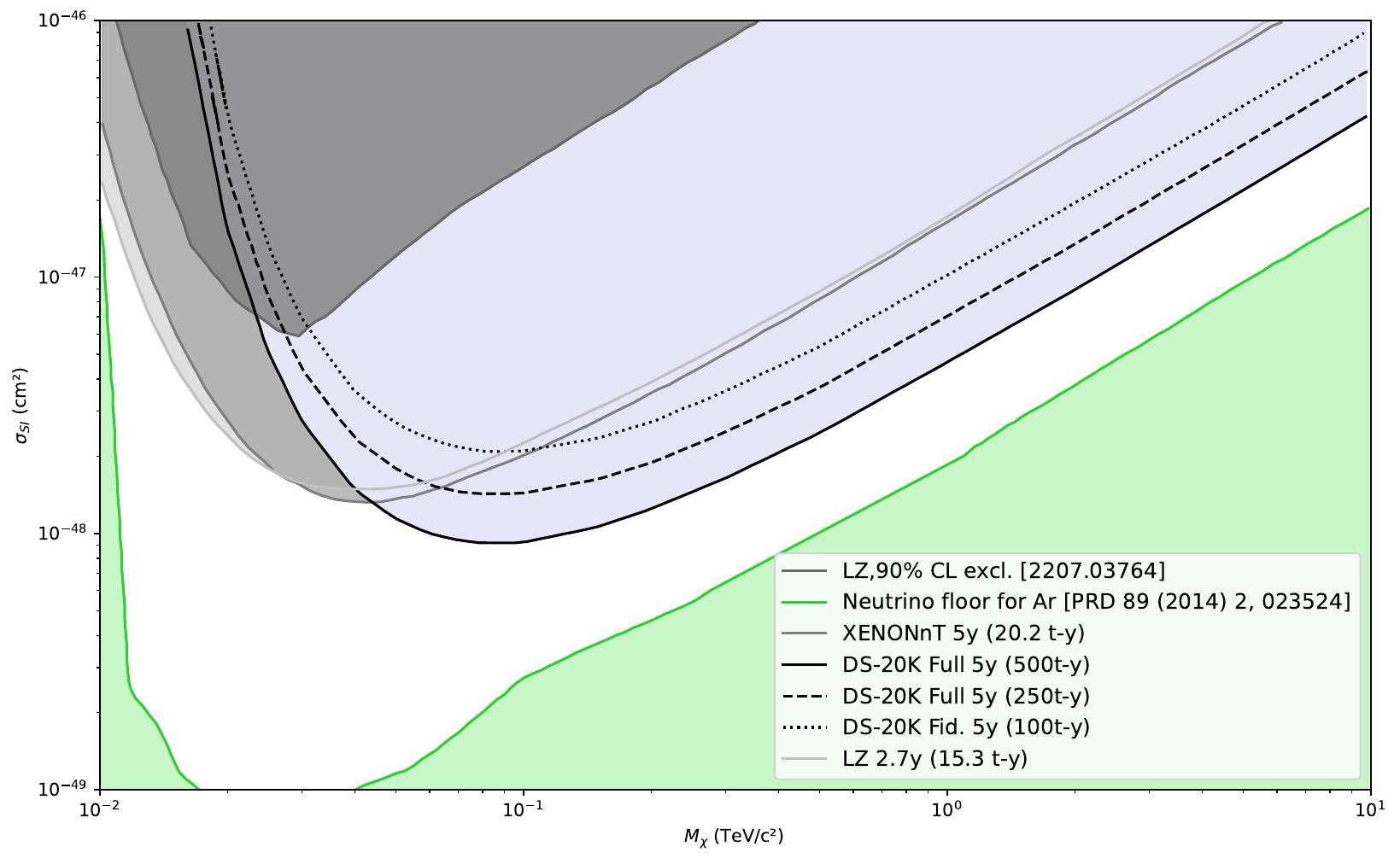}
    \caption{Current and Projected sensitivity of the DarkSide experiment on the dark matter--nucleon spin-independent scattering cross-section. Read the text for details.}
    \label{fig2}
\end{figure}

The DarkSide experiment \cite{DarkSide-20k:2017zyg} employs Liquid Argon (LAr) to search for dark matter particles through scintillation and ionization readout techniques \cite{DarkSide:2018kuk}. When dark matter interacts with Argon, a prompt scintillation light ($S_1$ signal) is produced, along with an ionization process. During ionization, electrons drift towards the anode of the time projection chamber, reaching a gas phase of Argon where they emit electroluminescence light ($S_2$ signal) \cite{DarkSide:2018bpj}. By leveraging the high trigger efficiency of the $S_2$ signal even for low-energy recoils \cite{DarkSide:2018bpj}, researchers can use the $S_2$ signal alone to constrain dark matter-electron scattering and establish restrictive bounds in the low dark matter mass regime \cite{DarkSide:2022knj,DarkSide-50:2022qzh} (see left-panel of FIG.~\ref{fig2}). In the previous white paper \cite{Abdalla:2019xka}, DarkSide was commissioning a 20-tonne LAr detector, but now we witness the construction of the DarkSide-20k at the Laboratori Nazionali del Gran Sasso (LNGS). It aims to search for WIMP Dark Matter particles scattering with liquid Argon nuclei in a 20t fiducial volume. It expects to reach a sensitivity to WIMP-nucleon cross-section of $7.4\times 10^{-48}\textrm{cm}^2$ for a TeV WIMP mass at the 90\% C.L with a 200 t.yr exposure \cite{Agnes:2023izm} (see right-panel of FIG.~\ref{fig2}).  We note that the DarkSide group has grown and consolidated at USP. As the traditional WIMP search for masses larger than $10$~GeV has excluded a large fraction of its region of interest, the USP group has contributed significantly on light dark matter probes (see left most plot on FIG.~\ref{fig2}). We highlight the best light dark matter constraints (albeit the small volume of DarkSide 50k) \cite{DarkSide:2018ppu}, bounds of the Migdal effect \cite{DarkSide:2022dhx}, and on leptophilic dark matter \cite{DarkSide:2018ppu}. Currently, they are improving the possibility for light dark matter detection with DarkSide 20k, and to the success of DarkSide 20k.  

\paragraph{CYGNO}

Conventional direct detection experiments will be challenged by the neutrino floor, and new avenues need to be explored, such as directionality \cite{Amaro:2023xhq,Amaro:2024prf,Amaro:2024vuk}. The CYGNO project plans to develop a high-precision optical Time Projection Chamber (TPC) for directional dark matter search and solar neutrino spectroscopy to be hosted at the LNGS. CYGNO's distinctive features include using scientific CMOS cameras and photomultiplier tubes coupled with a Gas Electron Multiplier for amplification within a helium-fluorine-based gas mixture at atmospheric pressure. The primary objective of this project is to achieve three-dimensional tracking with head-tail capability and to enhance background rejection down to the keV energy range. It is expected to yield competitive limits on the spin-independent dark matter-nucleon scattering cross-section for sub-GeV WIMPs and competing bounds on the spin-dependent scattering for GeV masses. We emphasize that CYGNOS represents a new addition compared to the last white paper.

\paragraph{OSCURA}

Oscura is a planned light-dark matter search experiment using Skipper-CCDs with a total active mass of 10~kg. It is based on 1.35 Mpix sensors packaged on a Multi-Chip-Module \cite{Oscura:2022vmi}. The ability to precisely measure the number of free electrons in each of the million pixels across a Skipper-CCD was demonstrated \cite{Tiffenberg:2017aac}, as well as the new physics reach to milicharged particles \cite{Oscura:2023qch}. 
The SENSEI Collaboration uses the same technology but features a total active mass of 100~g \cite{SENSEI:2023zdf}. The largest planned experiment using this technology is Oscura, which will certainly boost the sensitivity to light dark matter.

\section{Indirect Detection}

Indirect dark matter detection relies on observing the flux of stable particles such as electrons, protons, neutrinos, and gamma rays produced by dark matter annihilation or decay in dense astrophysical environments like dwarf spheroidal galaxies and the Galactic Center. For instance, the flux of gamma rays from dark matter annihilation is proportional to the dark matter annihilation cross-section, the energy spectrum (photons produced per annihilation), and the dark matter density along the line of sight, while being inversely proportional to the dark matter mass. Knowing the dark matter density we can estimate the flux of gamma rays from dark matter annihilation into a given final state. In the last white paper, there was Brazilian involvement in the CTAO, AMS-02, and H.E.S.S. telescopes. However, the Brazilian community has shifted its efforts to the Cherenkov Telescope Array Observatory (CTAO) and the Southern Wide-field Gamma-ray Observatory (SWGO) which are the main gamma-ray instruments in the next decades. It proves that the Brazilian community is dedicated to the science frontier. 

\paragraph{CTAO}

The CTAO will be the main observatory for high-energy gamma rays in the foreseeable future. The scientific agenda of CTAO is broad, and dark matter is one of them \cite{CTAConsortium:2017dvg,CherenkovTelescopeArray:2024osy}. With instrumental improvement, a wider field of view, and a large energy window covering gamma-rays from $20$~GeV to $300$~TeV, CTAO will surpass its predecessors in many ways, and it will be sensitive to dark matter interactions more than one order of magnitude better than current instruments. We highlight that the dark matter annihilation cross-section ($\sigma v = 3\times 10^{-26}\, {\rm cm}^3{\rm s}^{-1}$) within reach of CTAO is natural in several dark matter models \cite{Balazs:2017hxh}. The observatory will operate arrays on sites located in both the North and South hemispheres to provide full sky coverage and maximize the discovery potential for the rarest phenomena. In summary, CTAO has the potential to discover dark matter via the observation of high-energy gamma rays, and for this reason, CTAO is one of the flagship experiments. It is worth pointing out that Brazilian researchers have been actively working on dark matter studies in the scope of the CTAO.

\paragraph{SWGO}

\begin{figure}
    \centering
    \includegraphics[width=0.4\textwidth]{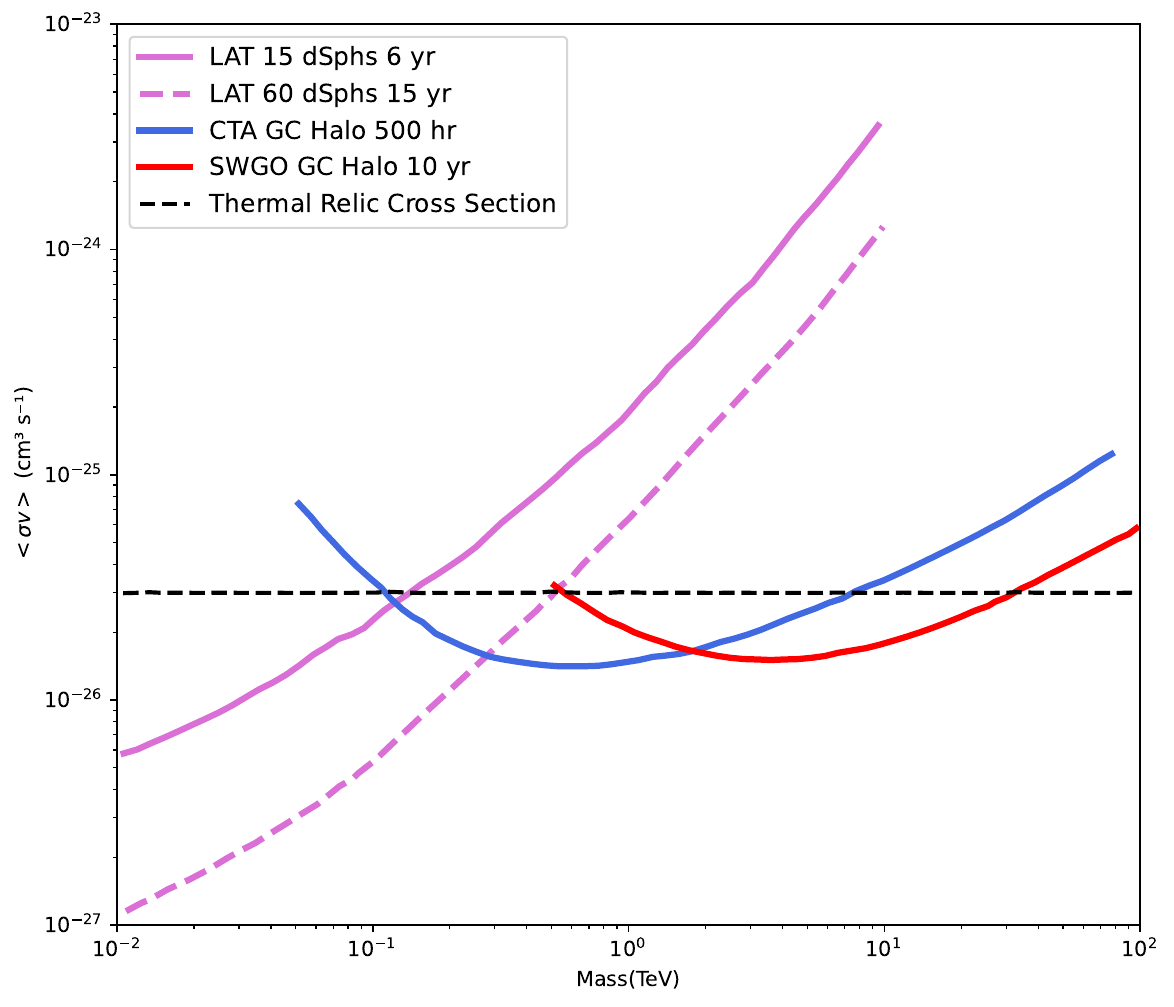}
    \includegraphics[width=0.4\textwidth]{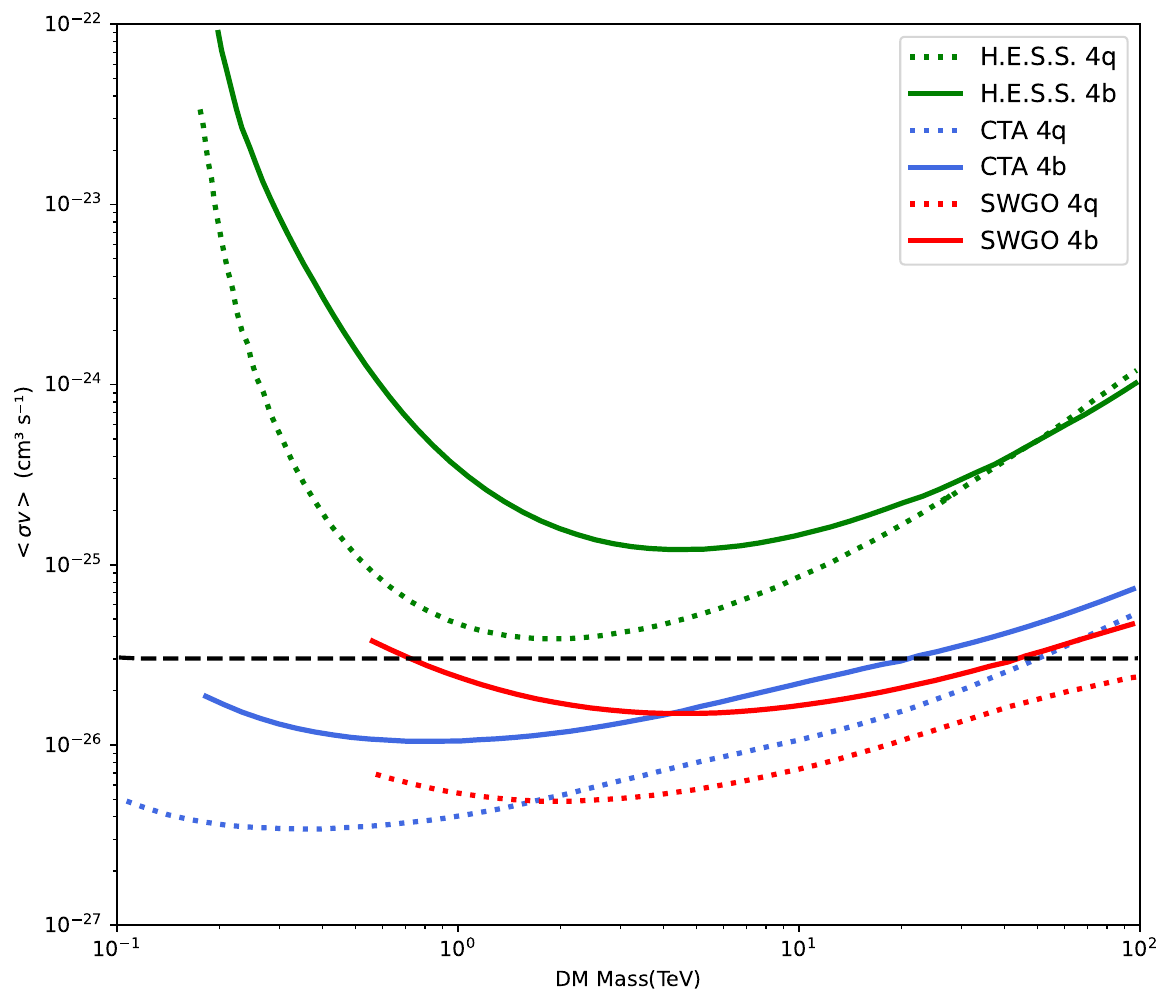}
    \caption{Projected limits on the dark matter annihilation into $b\bar{b}$ (left-panel) \cite{Viana:2019ucn} and into dark vector boson that eventually decay into quarks (right-panel) \cite{NFortes:2022dkj}.}
    \label{fig3}
\end{figure}

The SWGO is a proposed gamma-ray telescope that is fundamentally different from other Imaging Atmospheric Cherenkov Telescopes (IACTs) such as H.E.S.S. and CTAO. IACTs observe the Cherenkov radiation produced in the atmosphere by particle cascades initiated by incident high-energy gamma rays. SWGO, which is planned to be an Extensive Air Shower (EAS) array, will be able to collect data directly from particles in the shower that make it to the Earth’s surface continuously, even in the middle of the day. SWGO is designed to have an effective area and field of view larger than its cousin, The High-Altitude Water Cherenkov Observatory (HAWC) \cite{HAWC:2023owv}, but located in the Southern hemisphere. With all these features, SWGO will complement the CTAO sensitivity to dark matter, significantly beyond the 10~TeV dark matter mass region. Interestingly, several of the sensitivity studies have been carried out with the involvement of Brazilian researchers. It shows their active role in the science case of corresponding telescopes. In the left panel of FIG\ref{fig3}, we present the CTAO and SWGO sensitivity to dark matter annihilation into quarks \cite{Viana:2019ucn}, and in the right panel, their sensitivity to secluded dark matter \cite{NFortes:2022dkj}, where dark matter annihilates into dark vector bosons which later decay into quarks. It is visible from the plots that there is an interesting complementarity between the CTAO and SGWO telescopes in the TeV mass range.

\section{Colliders and Accelerators}

The Large Hadron Collider (LHC) probes the fundamental interactions through proton-proton collisions at high energies. Photons, charged and strongly interacting particles produce signals at the detectors and are used as triggers to search for dark matter particles. As dark matter particles interact weakly with matter and leave no trace at the detectors, their presence is inferred from energy-momentum reconstructions. In other words, dark matter particles are simply missing momentum at colliders. 

Generally speaking, missing momentum searches are dark matter searches, but there is also an effort to reinterpret other searches in terms of an extended ark sector, of which DM would be just a component. In this extended panorama, particle colliders play a crucial role in the search for dark sectors.
At colliders, dark sector searches typically involve pair production of dark matter particles, which yields a distinctive signature of missing transverse momentum. The community has developed simplified benchmark models to guide these searches~\cite{Abercrombie:2015wmb}, providing experimental targets and guidelines for comparing collider findings with those from direct and indirect detection experiments~\cite{Boveia:2016mrp,Albert:2017onk}. These benchmark models typically include a dark matter candidate and a mediator particle, which can also be a particle beyond the Standard Model spectrum. In the end, the ability of colliders to explore the properties of mediator particles and their interactions enhances our potential to uncover new physics, underscoring the importance of collider experiments in the broader search for dark sectors.

The design, deployment and monitoring of the trigger algorithms is a fundamental part of the LHC experiments, due to the high instantaneous luminosity delivered by the accelerator. Dark sectors pose challenges to the standard triggers and data acquisition setups~\cite{CMS:2024aqx}, which in turn impose the need for alternative strategies from the experiments~\cite{CMS:2024zhe}. 
As we do not know the properties of the dark matter particles, it is important to keep searching for signals with a wide array of detection strategies. That said, the Brazilian community is involved in dark sector studies within different experiments at the LHC, as described below.

\paragraph{ATLAS}

ATLAS (A Toroidal LHC ApparatuS) is a general purpose particle detector experiment at the Large Hadron Collider (LHC). The ATLAS detector has an inner tracking detector surrounded by a thin superconducting solenoid, electromagnetic and hadron calorimeters, and a muon spectrometer. The inner-detector system is subject to a $2$~T axial magnetic field that provides charged-particle tracking in the pseudorapidity range $|\eta| < 2.5 $. As it was designed to be a general-purpose experiment, ATLAS has been capable of probing several dark sectors \cite{ATLAS:2023kao,ATLAS:2023jyp,ATLAS:2024kpy,ATLAS:2024cju}. Interestingly, Machine learning (ML) methods have improved the event background
rejection. In particular, the Neural-Ringer algorithm (NR) \cite{DaFonsecaPinto:2777971,Spolidoro_Freund_2020} is currently being used in the community to probe vector dark matter particles.

\paragraph{CMS}

CMS (Compact Muon Solenoid) is also a multipurpose, nearly hermetic particle detector at the LHC. Similarly to ATLAS, it has a broad physics program and also features dedicated searches for dark sectors. The key feature of the CMS apparatus is a superconducting solenoid with a magnetic field of $3.8$\,T equipped with a silicon pixel and strip tracker. The detector also comprises a lead tungstate crystal electromagnetic calorimeter (ECAL), and a brass and scintillator hadron calorimeter (HCAL), each composed of a barrel and two endcap sections.
Finally, CMS is equipped with a muon spectrometry system embedded in the steel flux-return yoke outside the solenoid.

Besides the aforementioned searches for transverse momentum imbalance~\cite{CMS:2021far,CMS:2020ulv,CMS:2019zzl,CMS:2019ykj}, the CMS collaboration also reinterprets many of its other analyses in terms of searches for the dark sector.
Those include measurements of the Higgs boson invisible decay width~\cite{CMS:2023sdw},
which can constrain a posited H\,$\to$\,DM decays;
searches for dijet resonances~\cite{CMS:2019gwf}, 
reinterpreted as dark sector mediators;
and searches for long-lived particles that are present in many models that lead to DM candidates.
A comprehensive report on searches for the dark sector with the CMS experiment is available in Ref.~\cite{CMS:2024zqs}. 

\paragraph{LHCb}

The LHCb (Large Hadron Collider beauty) is an experiment dedicated to b-physics but has searched for dark sectors in the low mass regime. LHCb has been proven to be an effective laboratory to search for both prompt and displaced resonances especially if the particle has a substantial decay branching ratio to muons \cite{LHCb:2016inz,LHCb:2017trq,Rodriguez:2021urv}. In general, the LHCb targets low mass and short-lived particles because of its relatively low energy threshold and integrated luminosity compared to general purpose detectors (CMS and ATLAS) \cite{Ilten:2015hya,Ilten:2016tkc}. Interestingly, in 2018, LHCb implemented electron identification in the first high-level trigger which permitted LHCb to search for dark particles decaying into $e^+e^-$ \cite{Craik:2022riw}. In the next LHC run, LHCb will continue to fill the gaps in the search for dark matter covering dark mediators masses ranging from  $10^{-2}$~GeV to $50$~GeV \cite{Alimena:2019zri,Craik:2022riw}.

\paragraph{Future Colliders: HL-LHC and FCC}

Naturally, groups involved in the LHC detectors are also participating in detector upgrades needed for the High Luminosity-LHC (HL-LHC) that is planned to achieve an integrated luminosity of $\mathcal{L}=3$~ab$^{-1}$ operating with a center-of-mass energy of $14$~TeV \cite{CidVidal:2018eel}. We stress that the HL-LHC is a future accelerator that has been approved and it will be a reality in the upcoming years. In the process of building an $100$~TeV future accelerator as the Future Circular Collider (FCC), a Large Hadron electron Collider (LHeC) could be constructed \cite{LHeC:2020van} offering a complementary probe to dark sectors \cite{DOnofrio:2019dcp,Huang:2022ceu}. In the context of the FCC, there are two groups with signed MoU (Memorandum of Understanding) who are actively participating of the FCC feasibility study \cite{FCC:2018evy,FCC:2018vvp,FCC:2018byv,RebelloTeles:2023uig}. In summary, it is quite clear that the Brazilian community plays an important role in developing collider physics and is involved in science projects crucial to the search for dark matter.\\

Beyond traditional particle colliders, other high-intensity facilities can contribute to the search for dark sectors. A proposal to search for dark matter at the Brazilian synchrotron light source has been put forth \cite{Duarte:2022feb}. The European Spallation Source (ESS), a next-generation neutron source based on a high-power proton accelerator, also offers new opportunities to explore physics beyond the Standard Model.\\

\paragraph{LNLS}
Beyond traditional particle colliders, other high-intensity facilities can contribute to the search for dark sectors. A proposal to search for dark matter at the Brazilian synchrotron light source has been put forth \cite{Duarte:2022feb}. An idea to repurpose SIRIUS predecessor, UVX, showed that a 1-3 GeV positron beam impinging on a target, followed by a missing mass spectrum event reconstruction, could cover an unexplored region of parameter space of the dark photon model. A similar proposal has been done for Axion-like particles \cite{Angel:2023exb}. New experimental strategies to search for dark sectors are being explored without interfering with SIRIUS's original purpose as well. 

\paragraph{HIBEAM-NNBAR at ESS}

The ESS-driven experiment HIBEAM/NNBAR \cite{Abele:2022iml} is designed to investigate baryon number violation by searching for neutron conversions into sterile states or antineutrons. These processes could be mediated by hypothetical bosons associated with the dark sector and baryogenesis \cite{Burgman:2024rkm}, providing an alternative pathway to uncover new interactions and possible connections to dark matter. The Brazilian participation in the HIBEAM-NNBAR experiment \cite{Santoro:2024lvc} is focused on the development of the tracking detector and associated simulations. In particular, recent work has been devoted to the calorimeter fast-simulation and optimization of the Time Projection Chambers configuration, and simulations of the readout electronics. 


\section{Complementary Probes}

There are complementary probes that do not fall precisely into this direct, indirect, and collider search for dark matter. Nonetheless, they are capable of probing different properties of dark matter.

\subsection{Neutrino Physics}

\paragraph{CONNIE}


Silicon detectors have pushed the energy thresholds to lower values, allowing for higher sensitivity in the dark-matter low-mass range. CONNIE was the first experiment to use silicon charge-coupled devices
(CCD) at a nuclear reactor to look for coherent elastic
neutrino-nucleus scattering \cite{2019PhRvD.100i2005A,2022JHEP...05..017C} and to impose competitive constraints on BSM physics \cite{2020JHEP...04..054C}. CONNIE is located just outside the dome of the Angra 2 nuclear power plant in Brazil.
It is important to emphasize that CONNIE has a strong Brazilian involvement, representing over 35\% of its members, and is a largely Latin American effort in terms of overall participants, with key contributions from Fermilab. 

\begin{figure}
    \centering
    \includegraphics[width=0.5\linewidth]{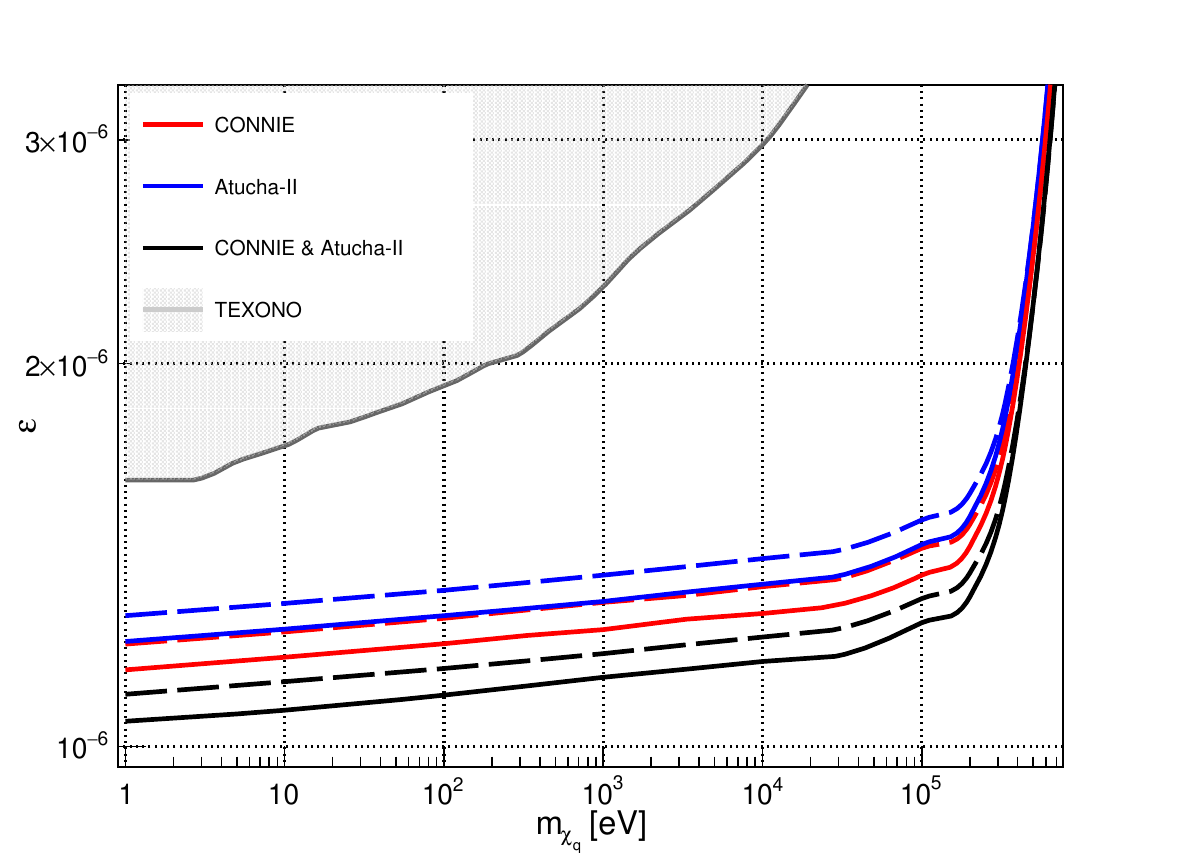}
    \caption{Upper limits on the charge of milicharged particles, $\chi$, at 90\% C.L., as a function of their mass. The red curve delimits the CONNIE bound. The blue one is from an experiment carried out at the Atucha power plant. The black line represents the limit from a combination of both datasets (see \cite{CONNIE:2024off}
    for details). The shaded region corresponds to the results obtained by the Texono experiment. Reproduced from \cite{CONNIE:2024off}
    } %
    \label{figconnie}
\end{figure}

Neutrino-nucleus coherent scattering (CE$\nu$NS) was firstly measured by COHERENT \cite{COHERENT:2020iec} in 2017 and now by XENONnT collaboration \cite{Adamski:2024yqt}. CONNIE aims to measure it at lower energies \cite{2019PhRvD.100i2005A,2022JHEP...05..017C,CONNIE:2024pwt} and consequently probe a different mass region of dark sectors. 
In 2021, CONNIE upgraded its detector by substituting the CCDs with a pair of Skipper-CCDs, employing for the first time these sensors for reactor neutrino detection, which increased its 
low-energy reach \cite{CONNIE:2024pwt}. Thanks to this, 
CONNIE has set world-leading limits on the charge of millicharged particles \cite{CONNIE:2024off}. 
Millicharged particles are fields beyond the Standard Model, which have a small electric charge $\epsilon e$, and interaction with photons of the type $\epsilon e \bar{\chi} \gamma_\mu \chi B_\mu$, where $B_\mu$ is the photon \cite{SENSEI:2023gie} (See Fig \ref{figconnie}).

In May 2024 CONNIE sensors were upgraded to host a 
Multi-Chip-Module (MCM), part of the R\&D for the Oscura \cite{Oscura:2022vmi} experiment (and first MCM operating in a working experiment), increasing its mass in skipper CCDs by a factor close to 30. With newly available multiplexing electronics, CONNIE can host several such modules. Therefore, we expect improved constraints on dark sectors. In addition, the collaboration aims to build a new detector to be installed inside the dome of the nuclear power plant, therefore increasing the neutrino flux and adding some overburden that will improve the shielding against cosmic rays.

\paragraph{SBND}

The SBND (Short-Baseline Neutrino Detector) presents an exciting opportunity to probe dark sectors. Light mediators that mix with photons can be produced in neutrino beams through meson decays and subsequently detected. Although similar to CONNIE, these neutrino detectors do not directly probe dark matter particles. They can provide complementary and crucial information about viable dark matter interactions within a dark sector and other new physics signals \cite{Machado:2019oxb,Alves:2024djc}. The signals from the dark sector are similar to those produced by neutrino coherent scattering, making neutrinos the primary background. However, spectral information and triggers on highly off-axis beams can help manage the neutrino background, enhancing the ability to detect dark sector interactions. This highlights the importance of neutrinos as a probe for dark sectors, offering a unique and valuable perspective in the quest to understand dark matter and its interactions \cite{Balasubramanian:2023pap,Acevedo:2024wmx}.

\paragraph{DUNE}

Deep Underground Neutrino Experiment (DUNE) will feature two neutrino detectors positioned within the world’s most intense neutrino beam. The first detector, located at the Fermi National Accelerator Laboratory in Batavia, Illinois, will capture particle interactions near the beam's source. The second, significantly larger detector, will be installed over a kilometer underground at the Sanford Underground Research Laboratory in Lead, South Dakota, situated 1,300 kilometers downstream from the source. 

The exploration and development of the novel technology of liquid-argon time-projection chambers (LArTPCs) present in large volume neutrino detectors such as DUNE. This allows us to probe some dark matter scenarios as boosted dark matter, which relies on the presence of a small and relativistic component of dark matter that can be detected via its interactions with the Standard Model particles \cite{Huang:2013xfa}. Signals from boosted dark matter that come in principle singled out from potential background sources (neutrinos) and thus make neutrino experiments dark sector detectors, as already happens with dark matter detectors \cite{DeRomeri:2019kic,Kelly:2019wow,DeRomeri:2021xgy,Acevedo:2024wmx}.

\subsection{Cosmology and Astrophysics Frontier}

\paragraph{DES}

The Dark Energy Survey (DES) is designed to probe the origin of the accelerating universe by measuring the 14-billion-year history of cosmic expansion with high precision. The collaboration is using an extremely sensitive 570-megapixel digital camera at Cerro Tololo Inter-American Observatory, high in the Chilean Andes. DES has been able to significantly improve the determination of the matter profile in galaxy clusters \cite{DES:2024rfx}, and investigate baryonic feedback using weak lensing combined with the kinetic Sunyaev Zel'dovich effect \cite{DES:2024iny}. Recently DES detected the presence of lensing above $5\sigma$ for the first time and consequently constrained the effective mass-to-light ratios and radial profiles of dark matter haloes surrounding individual galaxies \cite{DES:2024lto}. Hence, it is clear that DES continues to be an important probe of dark matter.

\paragraph{J-PAS}

Over the last decades, galaxy redshift surveys have achieved significant progress in studying the dark side of the universe dominated by non-baryonic dark matter and dark energy. The Javalambre Physics of the Accelerated Universe Astrophysical Survey (J-PAS) \cite{J-PAS:2014hgg} started this year by observing thousands of square degrees of the Northern sky using its unique set of 56 narrow-band filters covering the entire optical wavelength range. J-PAS will measure positions and redshifts for dozens of millions of galaxies and millions of Quasars, with an expected photometric redshift precision of $\sigma(z)\simeq0.003(1 + z)$, where $z$ is the redshift. Using different techniques, J-PAS data will be used to test the physical viability of different dark matter candidates and their possible interaction with other energy components and place tight constraints on the sum of the neutrino’s mass \cite{Salzano:2021zxk,Figueruelo:2021elm}.

\paragraph{BINGO}

Observations of the redshifted 21-cm line of neutral hydrogen (HI) open a powerful new window for mapping the spatial distribution of cosmic HI and advancing our knowledge of cosmology. BINGO (Baryon Acoustic Oscillations from Integrated Neutral Gas Observations) is a groundbreaking radio telescope designed to be one of the first to probe Baryon Acoustic Oscillations (BAO) at radio frequencies and through that bring complementary information on dark sectors \cite{Abdalla:2021nyj,Costa:2021jsk}. It is important to stress that dark sectors here refer to models that feature interactions between dark energy and dark matter components described by the equation of states. Nevertheless, BINGO represents a powerful window of observations that will improve our understanding of the $\Lambda$CDM model, in particular, the Hubble constant and the dark energy equation of state \cite{Novaes:2022ner,Costa:2021jsk}.

\paragraph{LSST}

Wide-field photometric surveys such as the Vera C. Rubin Observatory’s Legacy Survey of Space and Time (LSST) are instrumental for astrophysical probes of DM \cite{LSSTDarkMatterGroup:2019mwo,LSSTDarkEnergyScience:2021ryz,Sarcevic:2024tdr}. 
In particular, LSST will image 18,000 sq-deg to unprecedented depth in 6 filters, with over 800 visits in each spot along it's 10 year observation period, providing a ``movie'' of the sky.
LSST's image quality, area, depth, filters, time-domain observations and coverage including the galactic plane will enable a large swath DM tests, from the Milky-way to the high-redshift Universe.
LSST will observe Milky Way satellite galaxies, stellar streams, and strong lens systems, which are essential for detecting and characterizing the smallest dark matter halos. It will help us probe the minimum mass of ultra-light dark matter and thermal warm dark matter with unprecedented precision \cite{LSSTDarkMatterGroup:2019mwo}. Moreover, precise measurements of the density and shapes of dark matter halos in dwarf galaxies and galaxy clusters are crucial to search for dark matter signals in astrophysical objects \cite{Mao:2022fyx}.

\paragraph{LISA}
Gravitational wave (GW) detectors can also be used as direct or indirect probes of dark matter~\cite{Bertone:2019irm, LISA:2022kgy}. Among these experiments, the Brazilian community has a few members actively engaged in the Laser Interferometer Space Antenna (LISA) project. LISA is a planned space-based interferometer of triangular shape, with an arm length of $2.5$ million km, ensuring a sensitivity to waves in the mHz band. By the time of its launch in the mid-2030s it shall be the first GW detector to access this frequency range~\cite{LISACosmologyWorkingGroup:2022jok}. Interestingly, LISA will be sensitive to dark matter in the form of primordial black holes~\cite{Bird:2016dcv, Raidal:2017mfl}, light bosons \cite{Brito:2017zvb}, and provide orthogonal and complementary limits on WIMPs \cite{Arcadi:2023lwc}.

\paragraph{ET} The Einstein Telescope (ET) is a third-generation gravitational wave
detector that will cover a wide spectrum of frequencies measurable on the Earth’s
surface, in particular from a few Hz up to $10$~Hz \cite{Mangano:2024kve}. 
ET will need to develop innovative technologies to reduce background noise of thermal nature. For instance, new cooling techniques, cryogenic systems, and low mechanical dissipation materials \cite{Badaracco:2024kpm}. ET has recently received its first grant (2.6 millions euros) for research and development. It is a long-term project, but it is exciting that the community is paving the road for measuring Gravitational waves at these frequencies.
 
We will now transition from discussing the experimental endeavors to focusing on the theoretical efforts within the community. This shift will highlight how theories developed in the past are rigorously tested and refined in light of new data from the aforementioned experiments.

\section{Theoretical Efforts}

\paragraph{Dark Matter: Particle Physics}

The Brazilian groups working on dark sectors from a particle physics perspective have grown in the past years and interestingly their works are always in connection to experimental data from cosmology, collider physics and dark matter detectors. In an attempt to probe different dark matter properties, some works investigated the impact of heavy and unstable dark matter particles in extensive air shower detectors of gamma rays \cite{Esmaili:2021yaw} and on the IceCube neutrino detector~\cite{Bhattacharya:2019ucd}. Others studies focused on dark matter annihilation, either producing a continuous gamma-ray emission  \cite{CTA:2020qlo,NFortes:2022dkj,CTAConsortium:2023yak,Andrade:2024ekx} or gamma-ray lines \cite{Angel:2023rdd}.

In the context of direct detection, it is well known that light mediators may alter the neutrino floor as well as enhance the dark matter-nucleon scattering rate \cite{Lindner:2020kko}. The restrictive limits from direct detection experiments have excluded a large region of parameter space of several dark matter models assuming a thermal freeze-out production \cite{Arcadi:2017kky,Arcadi:2021yyr,Oliveira:2021gcw}. It has been shown that if dark matter particles have a sizeable non-thermal production the limits from direct experiments can be relaxed \cite{Arcadi:2020aot,Arcadi:2021doo,Barman:2021ifu,Dutra:2021lto}. As direct detection experiments face challenges to lower their energy threshold, it has been shown that neutron stars can fill the gap and probe light dark matter particles \cite{Maity:2021fxw}.

Direct and indirect constraints can also be avoided once we consider scenarios beyond the WIMP freeze-out paradigm.
In particular, models where dark matter is very weakly coupled to the SM through a heavy mediator can lead to interesting collider signatures. One interesting possibility is the conversion-driven freeze-out mechanism \cite{Garny:2017rxs,DAgnolo:2017dbv}, where the early freeze-out of dark matter leads to high co-annihilation rates with the mediator.
In these scenarios the parameter space region consistent with the observed relic abundance results in mediators with proper lifetimes of the order of cm to m. As a result, long-lived particle searches are sensitive to this model. It has been shown, however, that the current LLP search program at the LHC can be improved to test these models with higher sensitivity \cite{Heisig:2024xbh}.

Another interesting possibility is near-resonance production, which plays an important role in the proper computation of primordial relic density and where exclusion limits can be optimized for searches in proton-proton collisions at the LHC experiments and electron-position collisions at CLIC \cite{daSilveira:2023hmt,daSilveira:2024tpy}. In this regard, a proposal for dark sector detector has been put forth. It has been shown that the SIRIUS predecessor, UVX could be repurposed to directly produce dark photons \cite{Duarte:2022feb} or axion-like particles \cite{Angel:2023exb}, and become a competitive dark sector detector. Dark matter signals have been extensively searched for in proton-proton collisions. Having in mind, the proposal for a Future Hadron-Electron collider (LHeC), the sensitivity reach for leptophilic dark matter has been derived. It was shown that LHeC could probe WIMPS  \cite{Huang:2022ceu} as well as dark photons \cite{Oliveira:2022ypu}, and offer a complementary probe to the LHC. 

The interplay between dark matter and cosmology has also been explored. Interactions in the dark sector may alter the abundance of light elements \cite{Alves:2023jlo} and for this reason, one can use Big Bang Nucleosynthesis  as a laboratory for dark sectors  and particles. Moreover, non-thermal production of dark matter particles might help alleviate the intriguing $H_0$ tension \cite{Alcaniz:2022oow,deJesus:2022pux,daCosta:2023mow}. Dark matter particles which are thermally also leave imprints on cosmological observables. Having that in mind, constraints have been imposed on pseudoscalar dark matter using CMB data \cite{Rodrigues:2023xqu}.

Besides these works which are driven by experimental signatures, there were some on the model-building front. Experimental searches for axion and axion-like particles have set as a benchmark the KSVZ and DFSZ models \cite{Shifman:1979if,Zhitnitsky:1980tq,Dine:1981rt} \cite{Dias:2020kbj,Dias:2021lmf}.  These works interestingly show that the usual parameter space in terms of axion-photon coupling and axion mass can be modified, motivating the ongoing experimental effort to go search for signals beyond the KSVZ and DFSZ models. 

Part of the Brazilian community is also involved in designing machine learning algorithms that, for example, have been utilized to reconstruct dark states in high-energy collisions~\cite{Alves:2022gnw,Alves:2024sai} in an effort to characterize particles and interactions with essential information lost through neutrinos and/or dark matter.

\paragraph{Dark Matter: Cosmology and Astrophysics}

The Brazilian community devoted to cosmology and astrophysics research is very active, but most of the studies are not devoted to dark matter physics. Dark matter happens to be an interesting connection, with the main target being dark energy, modified gravity, and other cosmological observables.

For instance, many works investigated the connection between dark energy and dark matter equations of state in light of different datasets \cite{vonMarttens:2020apn,Benetti:2021div,Pereira:2022cmu, SantanaJunior:2024cug}. In other studies, the dark matter density was parameterized within bouncing cosmologies, where the field which dominates the contracting phase when the relevant CMB scales cross the Hubble radius must have the properties of dark matter \cite{Wands99,Peter:2006hx,Novello:2008ra,Pinto-Neto:2021gcl}. According to them, bouncing models have an advantage with respect to inflation because they require a field with the properties of an observed field, dark matter, instead of an inflation. Dark matter has also been parameterized in cosmological models in the presence of torsion \cite{Pereira:2022cmu}.

A more direct connection with dark matter physics has been investigated in the context of primordial black holes. Assuming a monochromatic mass function, one can derive constraints on the parameter space in which primordial black holes constitute a sizable fraction of the dark matter density using gamma-ray and neutrino data \cite{Capanema:2021hnm}. If primordial black holes obey an equation of state that has a tiny pressure component, their formation process changes \cite{Barroso:2024cgg}.  If such primordial black holes were formed in a dust-dominated contracting \cite{Magana:2022cwq} the formation process also changed, but this scenario does not lead to a significant mass fraction of primordial black holes as dark matter today \cite{Magana:2022cwq}. Extremal or remnant rotating black holes formed in the Planck energy have been studied in the context of dark matter as well. According to \cite{Carneiro:2021rwm,Borges:2023fub} such black holes could be produced at the reheating with the right dark matter abundance for sufficiently high reheating temperature. 

Regarding the dark matter density, there were investigations concerning the evolution law of the dark matter density \cite{Bora:2021uxq,Bora:2021iww,Santana:2023gkx} and estimations of the dark matter density using X-ray data from galaxy clusters \cite{Holanda:2019sod}. Other approaches based on machine learning algorithms trained on numerical simulations have been proposed, to make predictions on the total and dark matter profile of galaxies using observational survey \cite{vonMarttens:2021fmj,Wu:2023hwr}. The dark matter distribution in galaxies is usually satisfactorily described by a spherically symmetric density profile, but its density in the vicinity of type Ia supernovae requires attention \cite{Steigerwald:2021bro,Steigerwald:2021vgi,Steigerwald:2022pjo}. A new method to test dark matter profiles (or modified gravity) in large sample of galaxies was introduced in \cite{Rodrigues:2022oyd, Hernandez-Arboleda:2023abv}. A formation of a dark disk has also been investigated near the Sun \cite{Alves:2024qvj}. A proposal for understanding the dark matter profile of the Milky Way from analog galaxies was developed in \cite{deIsidio:2023osr}. In a different direction, the authors in \cite{Lima:2021slf} argued that the usual single power-law density profile is unsuitable for describing different lens mass intervals. 




\begin{figure}
    \centering
    \includegraphics[width=0.4\linewidth]{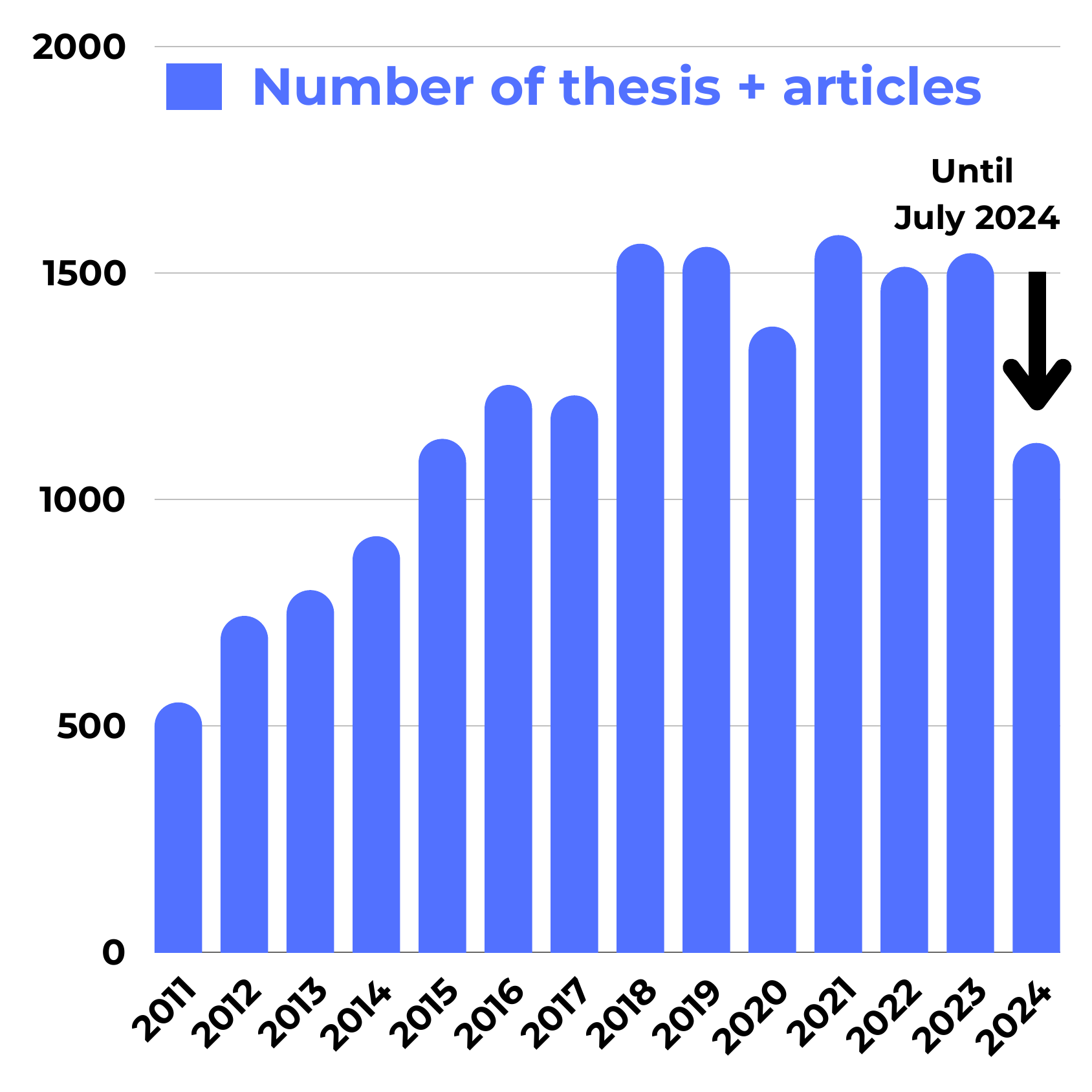}
     \includegraphics[width=0.4\textwidth]{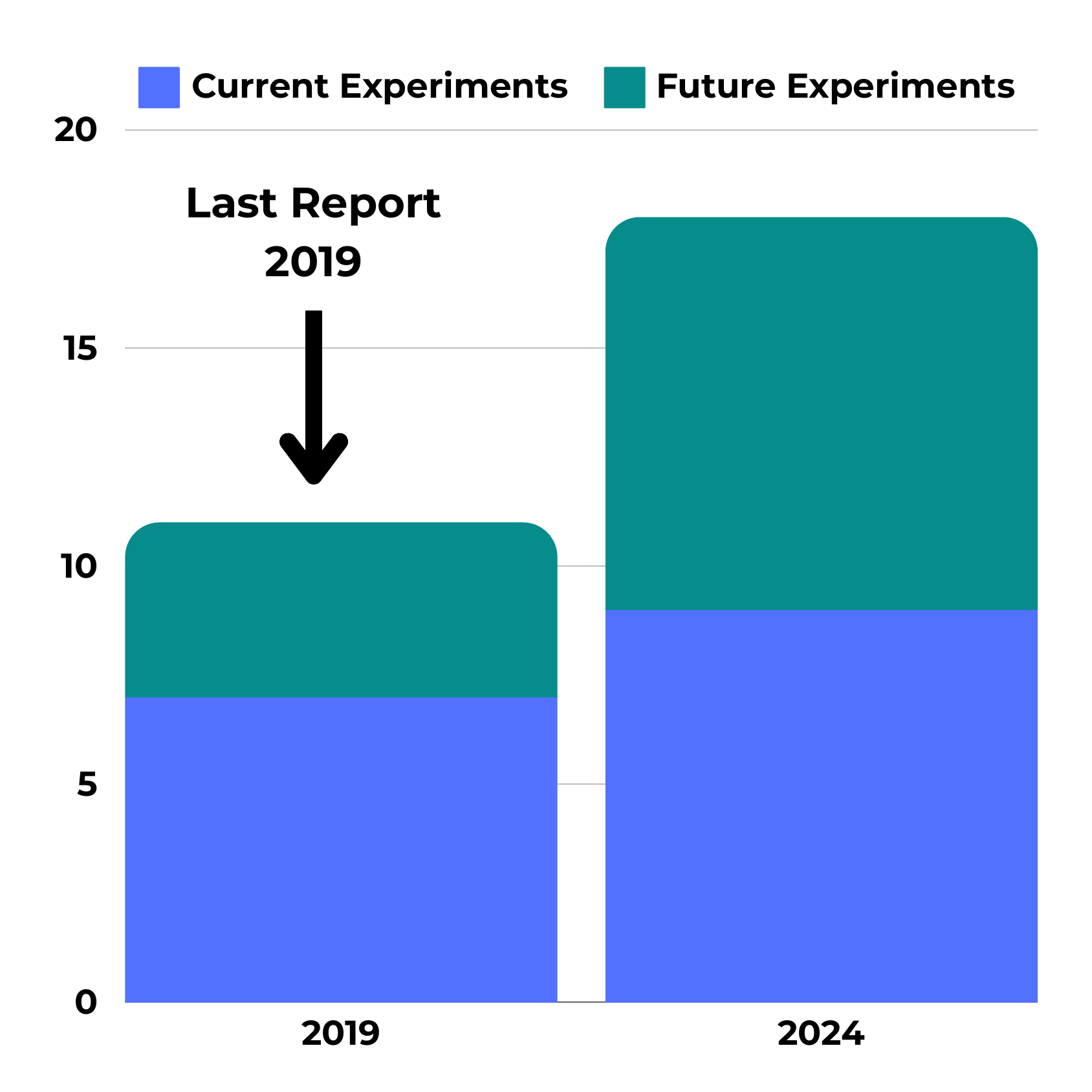}
    \caption{On the left, we show a summary plot with the number of works (thesis + papers) related to dark matter and dark sectors with a Brazilian affiliation. Data was collected using Google Scholar on July 22, 2024. In July 2024, we have produced more work than the whole year of 2014. On the right, we compare the number of experiments involved in the past white paper with this one. Our goal here is not to be rigorous about the precise number of experiments or works written, but to solidly show that the dark sector community is growing and aligned with a strong experimental program.}
    \label{fig:summary}
\end{figure}

\section{Conclusions}

It is important to put our conclusions into perspective. To the best of our knowledge, the first dark matter session at the National Meeting of Particles and Fields (now known as the Spring Meeting), promoted by the Brazilian Physical Society, took place in 2011. This indicates that the dark matter research community in Brazil is relatively young compared to others within the High Energy, Cosmology, Astrophysics, and Particles (HECAP) field. Despite its youth, the community has shown remarkable growth, as highlighted in the previous white paper back in 2019. This interest in dark matter physics can also be seen by the number of works presented in the last Spring Meetings. In FIG.~\ref{fig:dark_sectors} we show that the Brazilian community is experimentally involved in large-scale long-term science collaborations directly and indirectly related to dark matter physics. On the left panel, in FIG.~\ref{fig:summary}, we present an overview of all theses and articles, both published and unpublished, related to dark matter and dark sectors with Brazilian affiliations. The data, collected using Google Scholar on July 22, 2024, underscores the significant expansion and consolidation of Brazil's dark matter research community. While we do not aim for a precise count, our objective is to demonstrate that this community is active and aligned with international efforts in searching for dark matter and dark sectors. This progress reflects the increasing contributions and engagement of Brazilian researchers in this critical area of scientific inquiry.


\bibliographystyle{JHEPfixed}
\bibliography{references}
\end{document}